\documentclass[aps,reprint,showpacs,twocolumn,superscriptaddress,floatfix]{revtex4-1}
\usepackage{epsfig}
\usepackage[T1]{fontenc}
\usepackage[utf8]{inputenc}
\usepackage{amsmath}
\usepackage{amssymb}
\usepackage{amsfonts}
\usepackage{mathptmx}
\usepackage{cases}
\usepackage{textcomp}
\usepackage{dcolumn}
\usepackage{eucal}
\usepackage{bm}
\usepackage{color}
\usepackage[colorlinks,linkcolor=blue,citecolor=blue]{hyperref}

\usepackage{float}
\usepackage{epstopdf}
\DeclareMathAlphabet{\pazocal}{OMS}{zplm}{m}{n}
\begin{document}

	
	\title{Phase-tunable Josephson thermal router}

	\author{Giuliano Timossi}
	\email{giulianofrancesco.timossi@sns.it}
	\affiliation{NEST, Istituto Nanoscienze-CNR and Scuola Normale Superiore, Piazza S. Silvestro 12, I-56127 Pisa, Italy}

	\author{Antonio Fornieri}
	\thanks{Present address: Center for Quantum Devices and Station Q Copenhagen, Niels Bohr Institute,
		University of Copenhagen, Universitetsparken 5, 2100 Copenhagen, Denmark.}
	\affiliation{NEST, Istituto Nanoscienze-CNR and Scuola Normale Superiore, Piazza S. Silvestro 12, I-56127 Pisa, Italy}
	
	\author{Federico Paolucci}
	\affiliation{NEST, Istituto Nanoscienze-CNR and Scuola Normale Superiore, Piazza S. Silvestro 12, I-56127 Pisa, Italy}
		
	\author{Claudio Puglia}
	\affiliation{NEST, Istituto Nanoscienze-CNR and Scuola Normale Superiore, Piazza S. Silvestro 12, I-56127 Pisa, Italy}
	\affiliation{Dipartimento di Fisica dell'Università di Pisa, Largo Pontecorvo 3, I-56127 Pisa, Italy}
	
	\author{Francesco Giazotto}
	\email{francesco.giazotto@sns.it}
	\affiliation{NEST, Istituto Nanoscienze-CNR and Scuola Normale Superiore, Piazza S. Silvestro 12, I-56127 Pisa, Italy}
	
	
	\date{\today}



	
	\maketitle
	
	
	\textbf{
		Since the the first studies of thermodynamics, heat transport has been a crucial element for the understanding of any thermal system. Quantum mechanics has introduced new appealing ingredients for the manipulation of heat currents, such as the long-range coherence of the superconducting condensate
			~\cite{MakiGriffin,FornieriRev}. The latter has been exploited by phase-coherent caloritronics, a young field of nanoscience, to realize Josephson heat interferometers~\cite{GiazottoNature,MartinezNature,FornieriNature,FornieriNature2}, which can control electronic thermal currents as a function of the external magnetic flux.
		So far, only one output temperature has been modulated, while multi-terminal devices that allow to distribute the heat flux among different reservoirs are still missing. 
		Here, we report the experimental realization of a phase-tunable thermal router able to control the heat transferred between two terminals residing at different temperatures. Thanks to the Josephson effect, our structure allows to regulate the thermal gradient between the output electrodes until reaching its inversion. Together with interferometers~\cite{GiazottoNature,MartinezNature}, heat diodes~\cite{MartinezNatRect,MartinezAPL} and thermal memories~\cite{Guarcello2,FornieriPRB}, the thermal router represents a fundamental step towards the thermal conversion of non-linear electronic devices~\cite{FornieriRev}, and the realization of caloritronic logic components~\cite{PaolucciLogic,LiRev}.}
	
	\begin{figure}[t]
		\centering
		\includegraphics[width=0.42\textwidth]{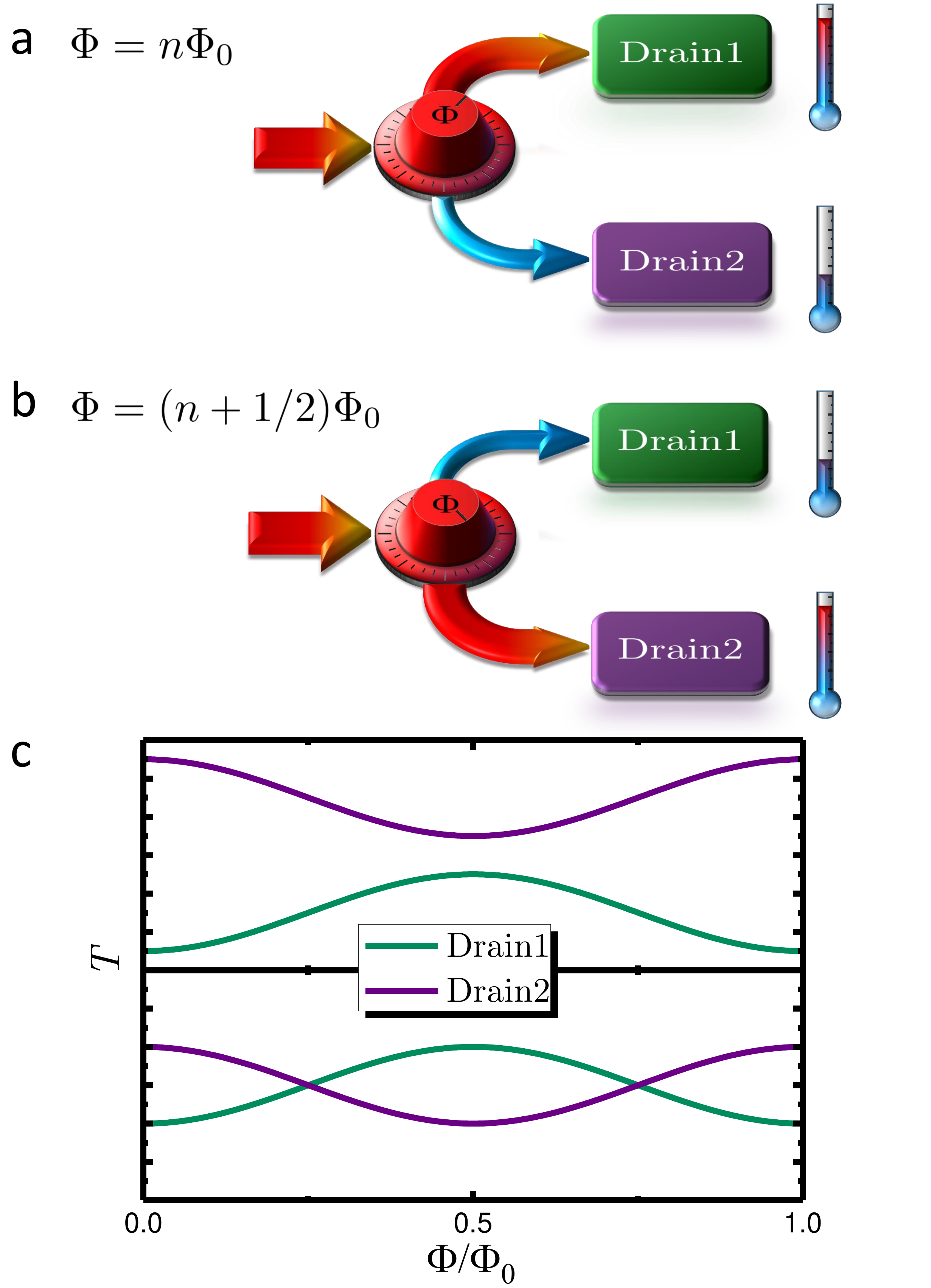}
		\caption{\textbf{Basic model of the Josephson thermal router.} \textbf{a.} Schematic representation of the phase-tunable Josephson thermal router. The heat current can be directed by varying the external magnetic flux. When $\Phi=0$ the heat current mainly flows to drain 1 and raises its temperature above that of drain 2. \textbf{b.} On the contrary, when $\Phi=\left(n+1/2\right)\Phi_0$ the heat current mainly flows to drain 2, where $n$ is an integer number and $\Phi_0$ is the flux quantum. This periodicity can be implemented by using a dc SQUID as routing element. \textbf{c.} Magnetic flux dependence of the drain temperatures. The router can work in two different regimes, depending on the bath temperature: in the \textit{splitting} regime, one of the output temperature is always higher than the other, while in the \textit{swapping} regime the temperature gradient can be inverted.
			\label{Fig1}
		}
	\end{figure}
	
	\begin{figure*}[t]
		\centering
		\includegraphics[width=0.85\textwidth]{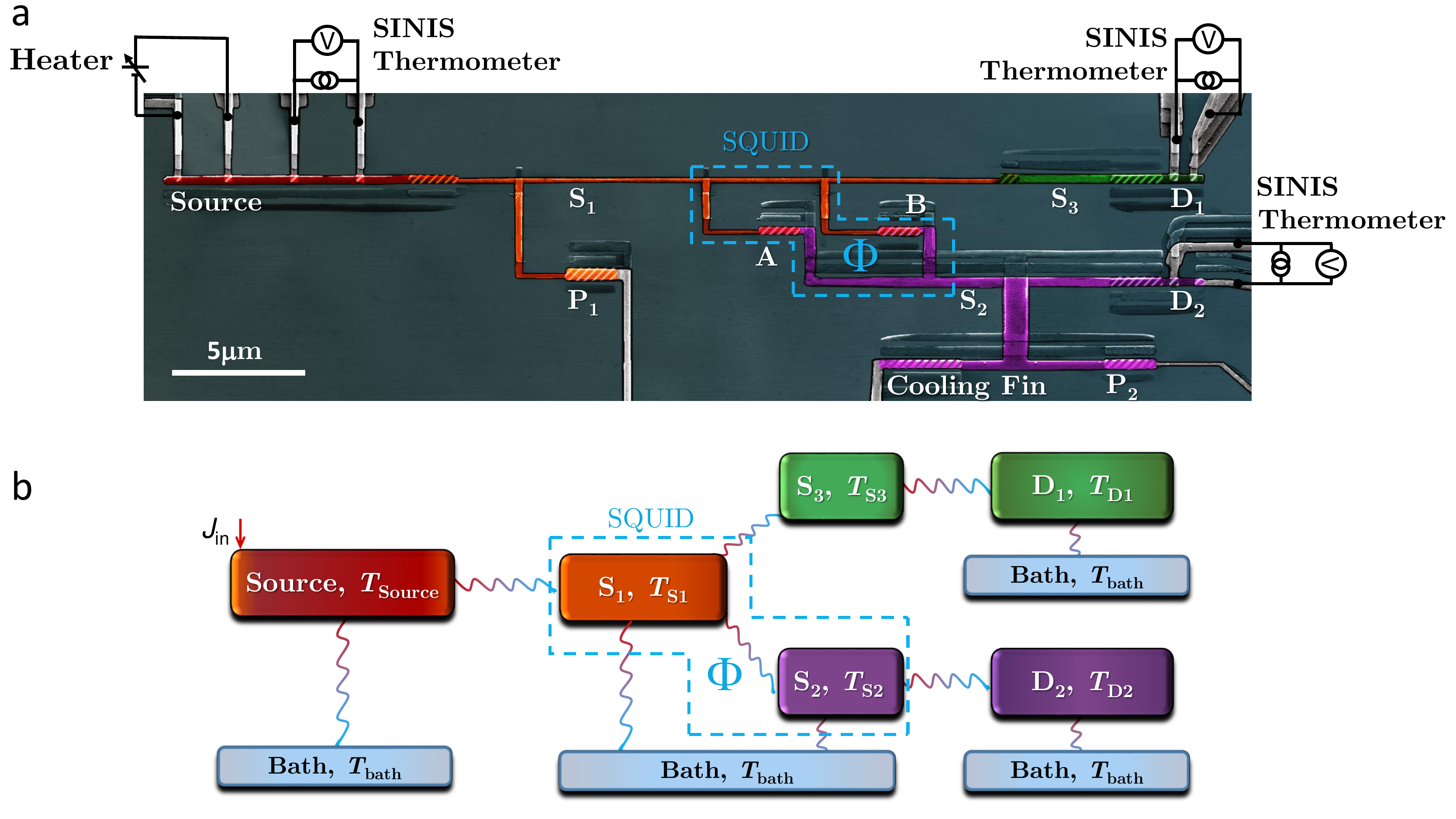}
		\caption{\textbf{Structure of the Josephson thermal router.} \textbf{a.} Pseudo-colour scanning electron micrograph of the device. Dashed regions indicate the junctions area: overlap of two materials with a layer of oxide among them. The source electrode (red), is coupled to four superconducting probes acting as thermometers and heaters. The source is also coupled with the upper branch of the SQUID S$_1$ (orange) from which the two superconducting islands with anti-phase temperature behaviour spread out, S$_3$ and S$_2$ (light green and light purple, respectively). Each island is connected to its respective normal metal electrodes D$_1$ (dark green) and D$_2$ (dark purple). A pair of superconducting probes for each drain is used to read the temperature. \textbf{b.} Thermal model describing the main heat exchange processes in the structure. Similar colours between the two images \textbf{a} and \textbf{b} have been chosen. Every spring represents a thermal interaction among the components. The colour gradient indicates the heat current direction from red to blue (if T$_{\rm source}$>T$_{\rm Bath}$). Here the knob (referred to Fig.~\ref{Fig1}a,b) is the core part between the source and drains. The entering heat flow in S$_1$ is divided between the two superconductors S$_3$ and S$_2$, and reaches at the end D$_1$ and D$_2$, respectively. 
			\label{Fig2}}
	\end{figure*}
	
	A thermal router is a device that allows to direct an incoming heat flow to a (drain) terminal of choice~\cite{Bosisio1} (see Fig.~\ref{Fig1}a, b). As shown in Fig.~\ref{Fig1}c, we can define two working regimes: in the \textit{splitting} regime, one of the drain temperatures is always higher than the other, while in the \textit{swapping} regime the output thermal gradient is inverted as a function of an external modulating parameter, i.e., the knob represented in Fig.~\ref{Fig1}a and~\ref{Fig1}b. This knob can be efficiently embodied by a magnetic flux $\Phi$ threading a temperature-biased direct-current superconducting quantum interference device (dc SQUID)~\cite{GiazottoNature}. The latter consists of a loop composed by two superconductors S$_1$ and S$_2$ residing at electronic temperatures $T_{\rm S1}$ and $T_{\rm S2}$, respectively, and connected by two Josephson tunnel junctions (JJs) in parallel, labelled `A' and `B'. If we impose a temperature gradient across the SQUID by raising $T_{\rm S1}$ above $T_{\rm S2}$, a steady state electronic heat current flows through the interferometer~\cite{GiazottoAPL}: 
	\begin{align}
	J_{\rm SQUID}(T_{\rm S1},T_{\rm S2},\Phi)&=J_{\rm qp}^{A}(T_{\rm S1},T_{\rm S2})\,(1+r) \notag\\
	&-J_{\rm int}^{A}(T_{\rm S1},T_{\rm S2})\,\sqrt{1+r^2+2r\mathrm{cos}\left( \frac{2\pi \Phi}{\Phi_0}\right) },\label{JSQUID}
	\end{align}
	where $J_{\rm qp}^{A}$ and $J_{\rm int}^{A}$ are the incoherent quasi particle contribution and the phase-coherent component of the heat current flowing through junction `A', respectively. The second term originates from energy-carrying tunneling processes involving creation and destruction of Cooper pairs on both sides of the JJs~\cite{Guttman,Barone} (see Methods). Consequently, it depends on the phase difference between the superconducting condensates across the JJs and can therefore be modulated by the magnetic flux piercing the SQUID loop~\cite{Tinkham}. Finally, in Eq.~\eqref{JSQUID} the parameter $r=R_{\rm A}/R_{\rm B}$ accounts for the asymmetry of the normal-state resistances $R_{\rm A}$ and $R_{\rm B}$ of the interferometer JJs, whereas $\Phi_0\simeq 2 \times 10^{-15}$ Wb is the superconducting flux quantum.\\
	\hspace*{0.42cm}Here, we experimentally demonstrate that a temperature-biased dc SQUID can work as the directional core of a Josephson thermal router able to control the thermal gradients between two drain electrodes in a superconducting hybrid circuit. The router operation is regulated by two main control parameters: the magnetic flux $\Phi$ threading the SQUID and the temperature of the lattice phonons. In our case, the latter are fully thermalized with the substrate phonons residing at the bath temperature, $T_{\rm bath}$, thanks to the vanishing Kapitza resistance between thin metallic films and the substrate at low temperatures~\cite{GiazottoRev,Wellstood,GiazottoNature,MartinezNature,MartinezNatRect,FornieriNature,FornieriNature2}. As we explained previously, $\Phi$ can modulate $J_{\rm int}$ according to Eq.~\eqref{JSQUID}, while instead the average value of the oscillating drain temperatures is mainly controlled by $T_{\rm bath}$. As we shall argue, by increasing the value of $T_{\rm bath}$, we can tune the working regime of the thermal router, which can operate as a splitter or as a swapper.\\
	\begin{figure*}[t]
		\centering
		\includegraphics[width=0.85\textwidth]{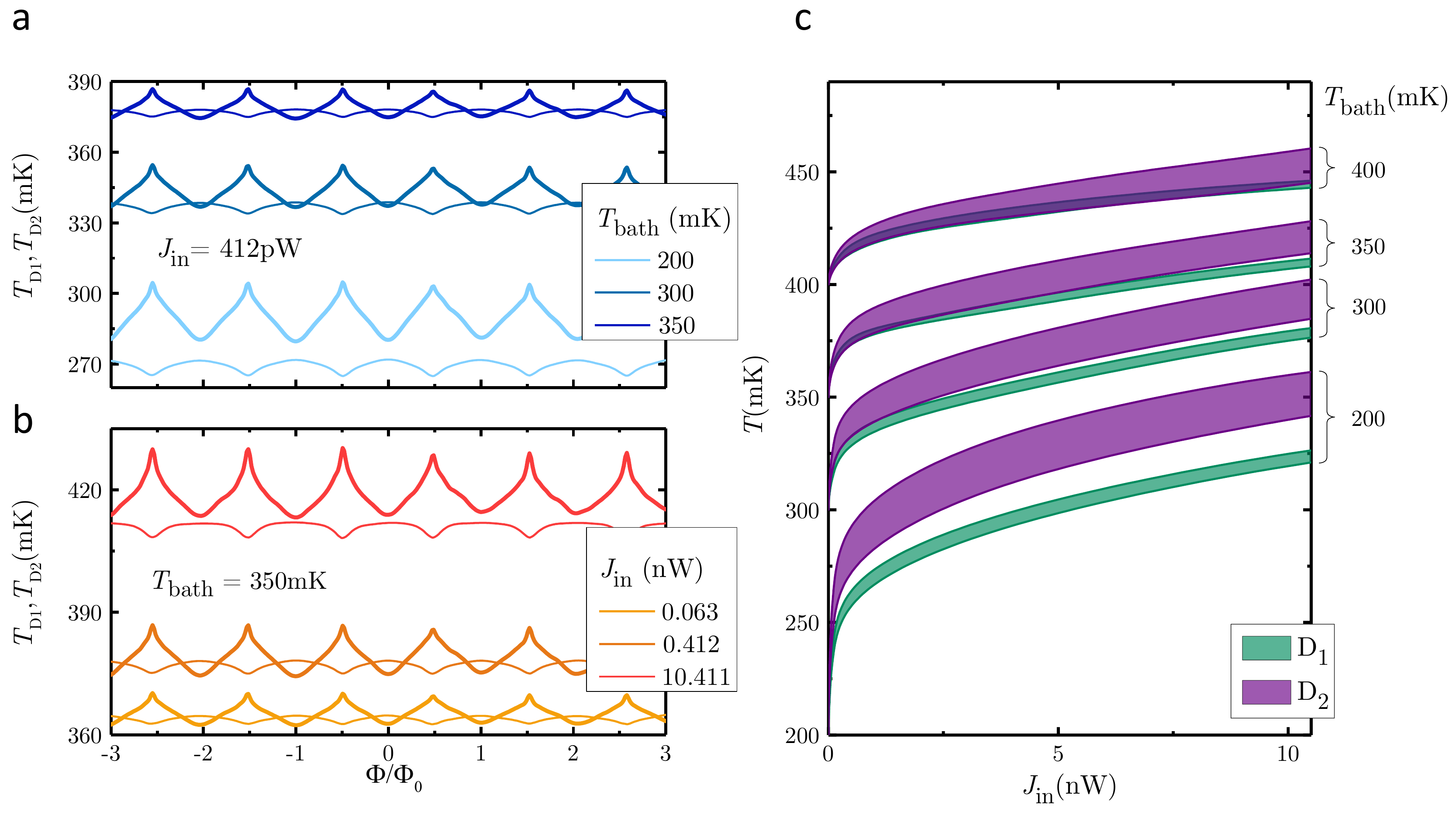}
		\caption{\textbf{Thermal behaviour of the router.} \textbf{a.} Drain temperatures $T_{\rm D1}$ (thick lines) and $T_{\rm D2}$ (thin lines) vs. the magnetic flux $\Phi$ piercing the loop of the SQUID for a fixed injected power $J_{\rm in}=412$ pW and different bath temperatures $T_{\rm bath}$. It is possible to distinguish between the \textit{splitting} regime (for $T_{\rm bath}\lesssim 300$ mK) and the \textit{swapping} regime (for $T_{\rm bath}\gtrsim 300$ mK). \textbf{b.} $T_{\rm D1}$ and $T_{\rm D2}$ vs. $\Phi$ for $T_{\rm bath}=350$ mK and different values of $J_{\rm in}$. \textbf{c.} Total swing of the drain temperatures as function of $J_{\rm in}$ for different values of $T_{\rm bath}$. The coloured regions indicate the amplitudes of the thermal oscillations at a given bath temperature. Four different values of $T_{\rm bath}$ are shown : 200 mK, 300 mK, 350 mK, and 400 mK.
			\label{Fig3}}
	\end{figure*}
	The realization of the thermal router is shown in Fig.~\ref{Fig2}a and~\ref{Fig2}b. The structure was fabricated by electron-beam lithography and three-angle shadow-mask evaporation of metals with \textit{in situ} oxidation (see Methods). Aluminium (with a critical temperature $T_{\rm c}\simeq1.55K$) was used for all the superconducting parts, and Al$_{0.98}$Mn$_{0.02}$ was used for the normal metal (N) electrodes~\cite{MartinezNature,MartinezNatRect,TaskinenAPL,FornieriNature}. The device consists of a N source electrode tunnel-coupled to the superconducting island S$_1$, forming the upper branch of the SQUID. As mentioned in the introduction, the electrode S$_1$ is connected to the lower branch of the SQUID S$_2$ by means of the JJs `A' and `B'. Moreover, S$_1$ is tunnel-coupled to the superconducting island S$_3$, which is necessary to reach the swapping regime, as we shall explain later. S$_3$ and S$_2$, in turn, are tunnel-coupled to the N drains D$_1$ and D$_2$, respectively (see Fig.~\ref{Fig2}a). 
	A N electrode acting as a cooling fin was tunnel coupled to S$_2$ (with a normal state resistance $R_{\rm F}\simeq1\ \rm k\Omega$) in order to maintain a suitable thermal gradient across the SQUID ~\cite{GiazottoRev,MartinezNatRect}. For the electrical characterization, two superconducting probes P$_1$ and P$_2$~\cite{Tinkham,GiazottoAPL} were tunnel-coupled to S$_1$ and S$_2$, respectively, with normal state resistances $R_{\rm P1}\simeq480\ \Omega$ and $R_{\rm P2}\simeq320\ \Omega$. The JJ between S$_1$ and S$_3$ ($R_{\rm S3}\simeq3\ \rm k\Omega$) was designed to obtain thermal decoupling of D$_1$ from S$_1$. The normal state resistances of the source and drains junctions are $R_{\rm Source}\simeq4.6$ k$\Omega$, $R_{\rm D1}\simeq1.5$ k$\Omega$ and $R_{\rm D2}\simeq1.5$ k$\Omega$, respectively. 
	Finally, source and drain electrodes are tunnel-coupled to superconducting probes (see Fig.~\ref{Fig2}a) realizing NIS junctions (normal state resistance of $\sim20\ \rm k\Omega$) used as thermometers and Joule heaters~\cite{GiazottoRev}.
	To extract the SQUID parameters, we measured the magnetic interference pattern of its Josephson critical current via a conventional four-wire technique, through P$_1$ and P$_2$ probes. From the fit of the data we obtained junction `A' normal-state resistance $R_{A}\simeq 650 \, \Omega$ and the asymmetry parameter $r \simeq 0.95$~\cite{GiazottoNature,Barone,SQUIDhandbook}.\\
	\hspace*{0.42cm}On the other hand, thermal measurements were performed using the set-up sketched in Fig.~\ref{Fig2}a. 
	By injecting a Joule power $J_{\rm in}$ into the source, we can raise its electronic temperature $T_{\rm Source}$ above $T_{\rm bath}$. In this way, we can increase the electronic temperature $T_{\rm S1}$ of S$_1$ and generate a thermal gradient across the SQUID at the core of the thermal router. This assumption is expected to hold thanks to the cooling fin connected to S$_2$. Indeed, the cooling fin extends into a large volume which provides a good thermalization with $T_{\rm bath}$ and represents an efficient channel to reduce $T_{\rm S2}$ below $T_{\rm S1}$.
	Therefore, a finite heat current $J_{\rm SQUID}$ flows through the SQUID and can be modulated by the magnetic flux $\Phi$ piercing its loop. In particular, if we set $T_{\rm Source}$ and $T_{\rm bath}$ to fixed values, $J_{\rm SQUID}$ is minimized when $\Phi=n\, \Phi_0$ (being $n$ an integer) [see Eq.~\eqref{JSQUID}] leading to a maximum of $T_{\rm S1}$. For $\Phi=n\, \Phi_0/2$, instead, $J_{\rm SQUID}$ is maximized and $T_{\rm S1}$ has a local minimum. As a consequence, the heat current flowing out of S$_1$ towards S$_3$ is maximized for $\Phi=n\, \Phi_0$, while the heat current flowing out of S$_2$ results to be minimized. The opposite situation is obtained when $\Phi=n\, \Phi_0/2$. Even though heat currents are not directly detectable, the described behaviour is reflected by the temperatures $T_{\rm D1}$ and $T_{\rm D2}$ of the drain electrodes connected to the different branches of the SQUID, which are expected to oscillate periodically with opposite phase dependence. Moreover, the superconducting island S$_3$ is introduced between S$_1$ and D$_1$ in order to reduce the difference between $T_{\rm D1}$ and $T_{\rm D2}$ and reach the swapping regime.\\
	\begin{figure*}[t]
		\centering
		\includegraphics[width=0.85\textwidth]{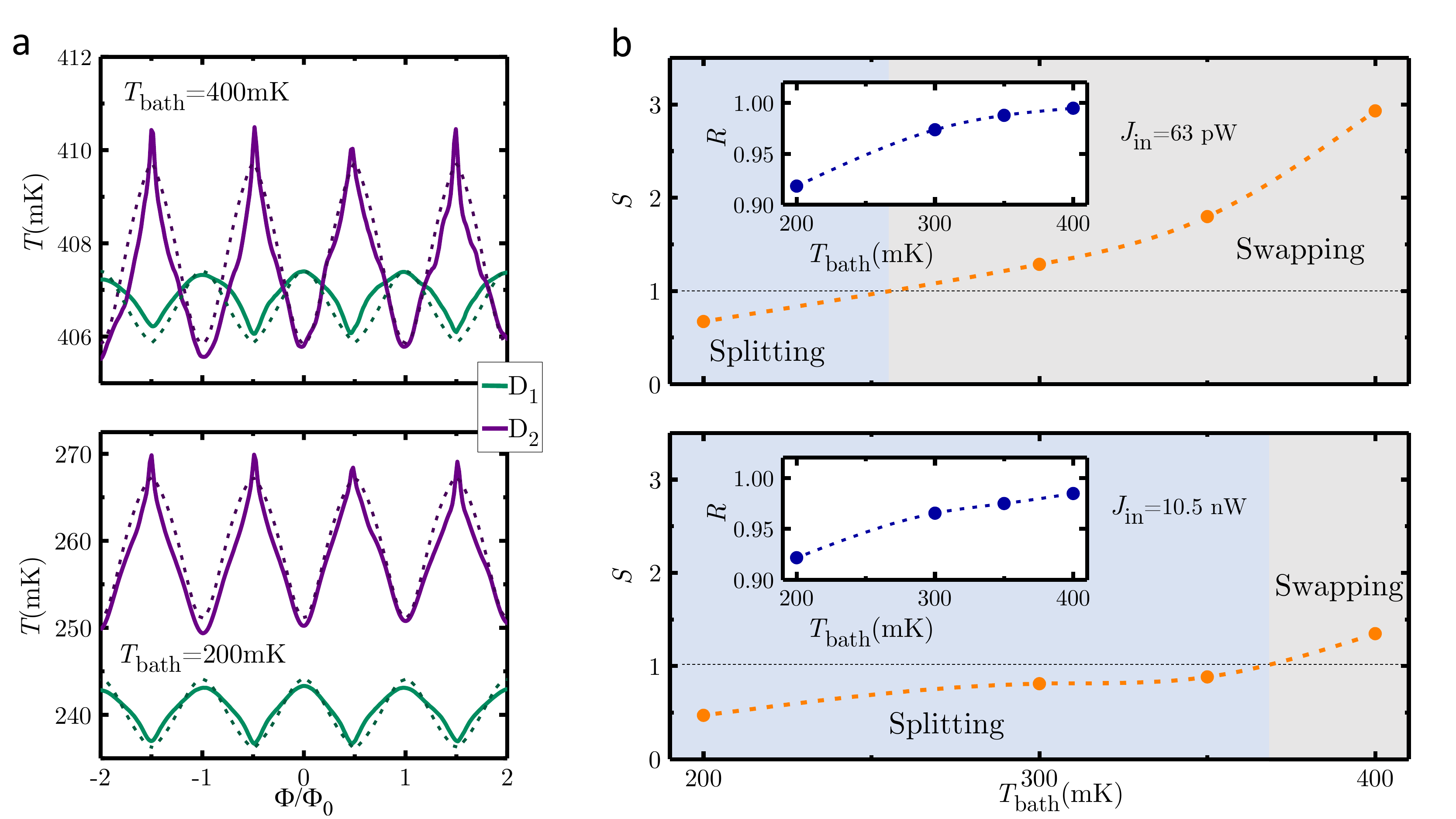}
		\caption{\textbf{Data fit and working regimes.} \textbf{a.} Magnetic flux modulations of $T_{\rm D1}$ and $T_{\rm D2}$ at two values of $T_{\rm bath}$ (200mK and 400mK) for $J_{\rm in}= 63$ pW. Solid lines represent the experimental data, while dashed lines are obtained from the thermal model of our structure (see Main Text). \textbf{b.} Performance parameter $S=\frac{\delta T_{\rm D1}+\delta T_{\rm D2}}{2 \left(\langle T_{\rm D2}\rangle - \langle T_{\rm D1}\rangle\right)}$ as a function of $T_{\rm bath}$ for $J_{\rm in}=63$ pW (upper panel) and $J_{\rm in}\simeq 10.5$ nW (lower panel). $S$ describes the working regime of the router, which operates as a \textit{splitter} (light blue area) for $S<1$ or as a \textit{swapper} (grey area) for $S>1$. The insets show the behaviour of $R=\langle T_{\rm D1}\rangle / \langle T_{\rm D2}\rangle$ as a function of $T_{\rm bath}$ for $J_{\rm in}=63$ pW and $J_{\rm in}\simeq 10.5$ nW. 
			\label{Fig4}}
	\end{figure*}
	The thermal behaviour of the D$_1$ and D$_2$ as a function of the magnetic flux is shown in Fig.~\ref{Fig3}a and~\ref{Fig3}b for different values of $T_{\rm bath}$ and $J_{\rm in}$, respectively. As anticipated, the drain temperatures manifestly demonstrate an opposite dependence on the magnetic flux: the minima and maxima of the curves corresponding to different drains are inverted. At low bath temperature, the oscillations do not overlap, so the thermal router is operating in the splitting regime. As we increase the value of $T_{\rm bath}$, the average temperatures of the drain electrodes $\langle T_{\rm D1}\rangle$ and $\langle T_{\rm D2}\rangle$ tend to approach each other, leading to overlapping modulations for $T_{\rm bath}\gtrsim 300$ mK. This behaviour is summarized by Fig.~\ref{Fig3}c, which displays the total swing amplitudes of $T_{\rm D1}$ and $T_{\rm D2}$ as a function of $J_{\rm in}$ for different bath temperatures. The injected power $J_{\rm in}$ also plays an important role, since it can raise $\langle T_{\rm D2}\rangle$ more efficiently than $\langle T_{\rm D1}\rangle$. This is due to the stronger coupling of S$_2$ to S$_1$ as compared to that between S$_1$ and S$_3$. In the splitting regime, we obtain a maximum separation between $T_{\rm D1}$ and $T_{\rm D2}$ of $\sim$40 mK for $J_{\rm in}\simeq10.4$ nW at $T_{\rm bath}=200$ mK, while in the swapping regime a maximum temperature inversion of $\sim$4 mK is obtained for $J_{\rm in}\simeq6.6$ nW at $T_{\rm bath}=400$ mK.
	To account for our observations we have elaborated a thermal model describing our structure (sketched in Fig.~\ref{Fig2}b). In this model, we take into account the predominant mechanisms of energy exchange, i.e, the electronic heat currents flowing through SIS, SIN or NIS junctions and the electron-phonon coupling in the normal metal electrodes (see Methods).
	For any given $T_{\rm Source}$, $T_{\rm bath}$ and $\Phi$, the steady-state electronic temperatures of every part of the device can be determined by solving the following system of energy-balance equations:
	\begin{subnumcases}{}
	-J_{\rm in}+J_{\rm Source}+J_{\rm Source,e-ph}=0\label{balanceSoruce}\\
	-J_{\rm Source}+J_{\rm P1}+J_{\rm SQUID}+J_{\rm S3}=0\label{balance1}\\
	-J_{\rm SQUID}+J_{\rm P2}+J_{\rm Fin}+J_{\rm D2}=0\label{balance2}\\
	-J_{\rm S3}+J_{\rm D1}=0\label{balance3}\\
	-J_{\rm D1}-J_{\rm Thermo}-J_{\rm 1}+J_{\rm D1,e-ph}=0\label{balanceD1}\\
	-J_{\rm D2}-J_{\rm Thermo}-J_{\rm 2}+J_{\rm D2,e-ph}=0\label{balanceD2}.
	\end{subnumcases}
These equations account for the detailed thermal budget of each electrode of the structure by equating the sum of all incoming and outgoing heat currents. In particular, Eq.~\eqref{balanceSoruce} refers to the thermal budget in the source, Eq.~\eqref{balance1} to S$_1$, Eq.~\eqref{balance2} to S$_2$, Eq.~\eqref{balance3} to S$_3$, and finally Eqs.~\eqref{balanceD1} and Eq.~\eqref{balanceD2} describe the energy exchange of D$_1$ and D$_2$, respectively. In the energy-balance equations, $J_{\rm P{1,2}}$ represent the heat currents released from S$_1$ and S$_2$ to the superconducting probes P$_1$ and P$_2$, $J_{\rm S3}$ is the heat flux between S$_1$ and S$_3$, whereas $J_{\rm Fin}$ accounts for the power delivered by S$_2$ to the cooling fin. Moreover, $J_{\rm D{1,2}}$ are the heat currents flowing into D$_{1,2}$, $J_{\rm {N,}\rm e-ph}$, where N=Source,D1,D2, are the heat losses due to the electron-phonon coupling in the electrodes and, finally, $J_{\rm Thermo}$ consider the energy released by D$_1$ and D$_2$ to the superconducting thermometers. Each term of the energy-balance equations is detailed in the Methods. In the model, we neglect photon-mediated thermal transport owing to poor impedance matching between the source and drains electrodes~\cite{meschke,Bosisio,Timofeev2,Pascal}, as well as electron-phonon coupling in all the superconducting parts of the structure due to their reduced volume and low  $T_{\rm bath}$~\cite{Timofeev1}.

The model provides a good agreement with the data for both drains at different bath temperature and different injected power, as shown in Fig.\ref{Fig4}a for $J_{\rm in}=63\rm pW$. Three parameters were extracted from the fit. The first is the resistance of the probe P$_2$ ($R_{\rm P2}=320 \Omega$), which is not directly obtainable from the electrical measurements (see Methods). The other two parameters are the parasitic heat currents $J_{1,2}$ injected by the thermometers of the drains, which remain on the order of the femtowatt (in particular J$_{\rm 1}$ varies from 1 fW to 8 fW and J$_{\rm 2}\sim-10$ fW). The origin of these parasitic powers may be thermal noise and fluctuations in the thermometers~\cite{meschke}.\\
In order to characterize the performance of our thermal router, we define two parameters outlining its working regime: the ratio between the average drain temperatures $R=\langle T_{\rm D1}\rangle / \langle T_{\rm D2}\rangle$, and the ratio between the sum of the amplitudes of the average drain temperature swings and the difference between the drain average temperatures $S=\frac{\delta T_{\rm D1}+\delta T_{\rm D2}}{2 \left(\langle T_{\rm D2}\rangle - \langle T_{\rm D1}\rangle\right)}$. As shown in the insets of Fig.~\ref{Fig4}b, $R$ displays the gradual approach of the drain average temperatures as $T_{\rm bath}$ is increased, which is not influenced by the injected power $J_{\rm in}$. On the other hand, $S$ takes into account the overlap of the thermal oscillations of the two drains. By definition $S$ is lower than 1 if the router works in the \textit{splitting} regime, while it is greater than 1 if the router is in the \textit{swapping} regime. The behaviour of $S$ as function of $T_{\rm bath}$ for two different values of $J_{\rm in}$ is shown in Fig.~\ref{Fig4}b. The injected power strongly affects $S$, influencing the transition point between the different working regimes of the thermal router. This is also clear from Fig.\ref{Fig3}c: at higher injected power the router reaches the swapping regime at larger values in $T_{\rm bath}$.

In summary, we have realized a phase-tunable Josephson thermal router able to control with very high accuracy the heat transferred among two terminals residing at different temperatures. The router provides high versatility of use, allowing to work in different regimes, easily accessible by varying the bath temperature. In the \textit{splitting} regime, we obtain a maximum separation of the average drain temperatures of $\sim$40 mK, whereas in the \textit{swapping} regime, our device can invert the thermal gradient between two reservoirs, with a maximum inversion of $\sim$4 mK. The latter capability may be also useful for mesoscopic thermal machines~\cite{Kosolof,Benenti}. The device is a proof of concept which can be easily improved by using a double loop SQUID~\cite{FornieriNature} to achieve a finer control, or a $0-\pi$ thermal JJ~\cite{FornieriNature2} to increase the amount of heat that can be controlled at the core of the structure. We wish to stress that our thermal router is a modular element, and it can be easily extended to multiple output terminals. Indeed the output of a thermal router can be used as the input for the next one, thus multiplying the output terminals. Finally, one of the fundamental requests of thermal logic~\cite{PaolucciLogic}, namely switching thermal signals among different channels, is answered by the phase-tunable thermal router.

\newpage
		\begin{center}
			{\large Acknowledgements}
		\end{center}
		We thank S. Gasparinetti for useful discussion. We acknowledge the MIUR-FIRB2013–Project Coca (grant no. RBFR1379UX), the European Research Council under the European Union’s Seventh Framework Programme (FP7/2007-2013)/ERC grant agreement no. 615187 - COMANCHE and the European Union (FP7/2007-2013)/REA grant agreement no. 630925 – COHEAT for partial financial support. The work of F.P. is funded by
		Tuscany Region under the FARFAS 2014 project SCIADRO.
		
		\begin{center}
			{\large Author contributions}
		\end{center}
		G.T. and F.P. fabricated the samples. G.T. and C.P. performed the measurements. G.T. and C.P. analysed the data and carried out the simulations with input from A.F. and F.G. G.T., A.F. and F.G. conceived the experiment. All the authors discussed the results and their implications equally at all stages, and wrote the manuscript.
		
		
		\clearpage
		\begin{center}
			{\large Methods}
		\end{center}
		\subsection*{Sample fabrication}
		The devices were fabricated with electron-beam lithography and three-angle shadow-mask evaporation of metals onto an oxidized Si wafer through a bilayer resist mask. The evaporations and oxidations were made using an ultra-high vacuum electron-beam evaporator. The latter allowed to deposit first 15 nm of Al at an angle of 30$^{\circ}$ to form the superconducting probes (thermometer and P$_1$), the lower branch of the SQUID and the S$_3$ part. Then the sample was exposed to 100 mTorr of O$_{2}$ for 5 minutes to realize the thin insulating layer of AlO$_{\rm x}$ forming the tunnel-barriers in all the SIS junction. Afterwards, 20 nm of Al were evaporated at 0$^{\circ}$ to deposit the superconducting island S$_1$ and the probe P$_2$. The insulating layers forming the tunnel barrier in all the SIN junctions were realized by another exposition to 400 mTorr of O$_{2}$ for 5 minutes. Finally, the sample was tilted at an angle of 30$^{\circ}$ and a deposition of 25 nm of Al$_{0.98}$Mn$_{0.02}$ was performed to implement the electrodes (source, D$_1$ and D$_2$) and the cooling fin. The source has a volume $V_{\rm Source}=1.1\times10^{-19}m^{-3}$, while the two drains have a volume $V_{\rm D1,2}=1.5\times10^{-20}m^{-3}$.
		
		\subsection*{Measurements}
		All measurements were performed down to 25 mK in a filtered dilution refrigerator. The SINIS thermometers were current biased by means of a low-noise floating source, while the voltage drop across the junctions was monitored with a standard room-temperature preamplifier. All the values of the temperature were extracted from an average of at least 10 measurements. The heaters were piloted with voltage biasing in the range 0-40 mV, corresponding to a maximum Joule power of $\sim 10$ nW injected in the source electrode. 
		
		\subsection*{Electrical-Thermal model}
		The electrical characterization data at different $T_{\rm bath}$ were fitted with a model including a finite screening parameter $\beta_K$~\cite{SQUIDhandbook}, due to a non-negligible value of the ring inductance~\cite{Guarcello}. From the fit we extracted $\beta_K=0.4$  and $T_{\rm c}=1.55\rm K$. However, the impact of the screening parameter on the thermal behaviour of our device results to be negligible: the hysteric effect for $\beta_K=0.4$ is at most $\sim$300$\rm \mu K$ for our data.
		
		In our thermal model, we considered the predominant energy exchange mechanisms in our system. First, the electronic thermal current flowing through NIS junctions, $J_{\rm NIS}(T_N,T_S)=(2/e^2R_i)\int_{0}^{\infty}\varepsilon\pazocal{N}\left(\varepsilon,T_S\right)\left[f\left(\varepsilon,T_N\right)-f\left(\varepsilon,T_S\right)\right]d\varepsilon$, where $f\left(\varepsilon,T\right)=\left[1+\exp\left(\varepsilon/k_BT\right)\right]^{-1}$ is the Fermi-Dirac distribution, $\pazocal{N}\left(\varepsilon,T\right)=\left|\Re\left[\left(\varepsilon+i\Gamma\right)/\sqrt{\left(\varepsilon+i\Gamma\right)^2-\Delta^2\left(T\right)}\right]\right|$ the smeared normalized Bardeen-Cooper-Schrieffer density of states (BCS DOS) in the superconductor~\cite{ Dynes}, $\Delta(T)$ is the temperature-dependent energy gap~\cite{Tinkham}, $\Gamma$ is the phenomenological broadening parameter known as gamma of Dynes~\cite{Dynes}, $R_i$ the tunnel junction normal state resistance, $e$ the electron charge and $k_B$ the Boltzmann constant. In our structure, we have $J_{\rm Source}\left(T_{\rm Source},T_{1}\right)=J_{\rm NIS}\left(T_{\rm Source},T_1\right)$, $J_{\rm D{1,2}}\left(T_{3,2},T_{\rm D{1,2}}\right)=-J_{\rm NIS}\left(T_{\rm D{1,2}},T_{\rm 3,2}\right)$ and $J_{\rm Fin}\left(T_{2},T_{\rm bath}\right)=-J_{\rm NIS}\left(T_{\rm bath},T_{\rm 2}\right)$. Since the thermometers connected to the drain electrodes form SINIS junctions~\cite{GiazottoRev}, $J_{\rm Thermo}(T_N,T_S,V)=(2/e^2R_i)\int_{-\infty}^{\infty}\left(\varepsilon-eV\right)\pazocal{N}\left(\varepsilon,T_S\right)\left[f\left(\varepsilon,T_S\right)-f\left(\varepsilon-eV,T_N\right)\right]d\varepsilon$.
		On the other hand, two components appear in the expression for $J_{\rm SQUID}$. The first accounts for the heat carried by quasiparticles, $J_{\rm qp}(T_{1},T_{2})=(2/e^{2}R_{\rm j}) \displaystyle\int_{0}^{\infty} \mathrm{d}\varepsilon \, \varepsilon \pazocal{N}_1(\varepsilon , T_{\rm 1})\pazocal{N}_2(\varepsilon , T_{2}) \allowbreak [f(\varepsilon ,T_{\rm 1})-f(\varepsilon , T_{2})]$, where $\pazocal{N}_1$ and $\pazocal{N}_2$ are the normalized BCS DOS of the superconductors forming the JJs. The second component represents the phase-coherent part of the heat current~\cite{,MakiGriffin,Guttman}, which reads $J_{\rm int}(T_{1},T_{2})=(2/e^{2}R_{\rm j}) \displaystyle\int_{0}^{\infty} \mathrm{d}\varepsilon \, \varepsilon \pazocal{M}_1(\varepsilon , T_{\rm 1})\pazocal{M}_2(\varepsilon , T_{2}) \allowbreak [f(\varepsilon ,T_{\rm 1})-f(\varepsilon , T_{2})]$. Here $\pazocal{M}_{\rm n}(\varepsilon,T) =\allowbreak \left|\Im\left[-i \Delta_{\rm n}(T)/\sqrt{\left(\varepsilon +i\Gamma_{\rm n}\right)^2-\Delta_{\rm n}^2(T)}\right]\right|$ is the Cooper pair BCS DOS in the $n$th superconductor, with $n=1,2$. In the steady state, the total heat current flowing through a Josphson junction can be expressed as $J_{\rm S1S2}(T_1,T_{2},\varphi)=J_{\rm qp}(T_1,T_{2})-J_{\rm int}(T_1,T_{2})\rm\; \cos\left( \varphi \right)$, where $\varphi$ is the phase difference between the two superconducting condensate. In the same way, $J_{\rm P{1,2}}=J_{\rm qp}(T_{1,2},T_{\rm bath})-J_{\rm int}(T_{1,2},T_{\rm bath})$ and $J_{\rm S3}=J_{\rm qp}(T_{3},T_{\rm bath})-J_{\rm int}(T_{3},T_{\rm bath})$. Finally, the electron-phonon coupling in the normal metal electrodes generates the term $J_{\rm N,e-ph}\left(T_{\rm N},T_{\rm bath}\right)=\Sigma \mathcal{V}_{\rm N}\left(T^6_{\rm N}-T^6_{\rm bath}\right)$, where N=Source,D1,D2 and $\Sigma=4.5\time10^9 \rm WK^-6m^-3$ is the material-dependent electron phonon coupling constant for Al$_{0.98}$Mn$_{0.02}$ ~\cite{MartinezNature,MartinezNatRect,TaskinenAPL}.\\
		The theoretical curves in Fig.\ref{Fig4}b,c were calculated from the thermal model using the measured values of $R_{\rm Source}$, $R_{\rm D1}$, $R_{\rm D2}$,$R_{\rm S3}$, $R_{\rm P1}$, $R_{\rm SQUID}$, $r$, $\beta_K$, $T_{\rm c}$ as determined from the electrical characterization of the devices. The experimental data were fitted by introducing the two parasitic injected powers in the two drains. These are about 10 fW and, so, compatible with residual heat input due to coupling with the environment. The energy balance equation of the source (Eq.~\ref{balanceSoruce}) does not take in account the heat losses by thermometer and involuntary heat current like $J_{\rm 1,2}$ since these terms are estimated to be at least two order of magnitude smaller.\\
		Furthermore, we had estimated the impact on the thermal behaviour of the non-negligible value of the ring inductance~\cite{Guarcello}. This makes rise of a finite $\beta_K$, but its impact on the device thermal behaviour turns out to be too small to be appreciated.\\
		The heat exchange due to the quasiparticle-phonon coupling~\cite{Timofeev1} in all the superconductors was also neglected, because this contribution turns out to be two orders of magnitude smaller than all the other thermal current in the explored range of temperatures.\\
		
		\newpage

\end{document}